\documentclass[aps,prl,twocolumn,superscriptaddress]{revtex4-1}
\usepackage{amsmath}
\usepackage{amssymb}
\usepackage{graphicx}
\graphicspath{{../figures/}}
\usepackage{subfigure}
\usepackage{verbatim}
\usepackage{amsfonts}
\usepackage{dcolumn}
\usepackage{bm}
\usepackage{color}
\usepackage[colorlinks,citecolor=blue]{hyperref}

\usepackage{mathrsfs}

\newcommand{\beq}{\begin{eqnarray} }
\newcommand{\eeq}{\end{eqnarray} }
\newcommand{\Beq}{\begin{eqnarray*} }
\newcommand{\Eeq}{\end{eqnarray*} }
\newcommand{\Bmat}{\left(\begin{matrix}}
\newcommand{\Emat}{\end{matrix}\right)}
\newcommand{\up}{\uparrow}
\newcommand{\dn}{\downarrow}

\begin{document}

\title{Thermal Properties and Instability of a $U(1)$ Spin Liquid on the Triangular Lattice}

\author{Qi-Rong Zhao}
\affiliation{Department of Physics, Renmin University of China, Beijing 100872, China}

\author{Zheng-Xin Liu}
\email{liuzxphys@ruc.edu.cn}
\affiliation{Department of Physics, Renmin University of China, Beijing 100872, China}
\affiliation{Tsung-Dao Lee Institute \& School of Physics and Astronomy, Shanghai Jiao Tong University, Shanghai  200240, China} 

\date{\today}

\begin{abstract}

We study the effect of Dzyaloshinskii-Moriya (DM) interaction on the triangular lattice $U(1)$ quantum spin liquid (QSL) which is stabilized by ring-exchange interactions. A weak DM interaction introduces a staggered flux to the $U(1)$ QSL state and changes the density of states at the spinon fermi surface. If the DM vector contains in-plane components, then the spinons gain nonzero Berry phase. The resultant thermal conductances $\kappa_{xx}$ and $\kappa_{xy}$ qualitatively agree with the experimental results on the material EtMe$_3$Sb[Pd(dmit)$_2]_2$.  Furthermore, owing to perfect nesting of the fermi surface, a spin density wave state is triggered by larger DM interactions. On the other hand, when the ring-exchange interaction decreases, another anti-ferromagnetic (AFM) phase with $120^\circ$ order shows up which is proximate to a $U(1)$ Dirac QSL. We discuss the difference of the two AFM phases from their static structure factors and excitation spectra. 


\end{abstract}

\maketitle
{\it Introduction.} Quantum spin liquids (QSLs)\cite{Anderson1987}, being exotic phases of matter, do not contain conventional long-range order even at zero temperature. Instead, QSLs support fractional excitations \cite{WenPSG} and have nontrivial many-body entanglement  \cite{ChenGuWen2010} which are beyond the Landau-Gintzburg paradigm. The anyonic excitations in QSLs can be applied in topological quantum computations\cite{Kitaev03, Anyon_TQC08} and have attracted lots of research interest. Antiferromagnetic spin systems may fall into quantum spin liquid phases when quantum fluctuations are strong enough such that long-range magnetic orders are melt. The quantum fluctuations can be enhanced by geometric frustration \cite{Anderson1987} and competing interactions\cite{Kitaev06}. Exactly solvable models, including the Kitaev model on honeycomb lattice\cite{Kitaev06} whose ground states are disordered QSL states,  have been constructed and provide guidelines for experimental realizations of QSLs. Candidate spin liquid materials have been discovered or synthesized, including antiferromagnets in the triangular lattices\cite{ETsalt03, dmit-science, YbMgGaOZhangQM, ExpU1Dirac_NP19}, the kagome lattices\cite{HebertsmithiteHanTH}, the honeycomb lattice \cite{RuCl3, Na2IrO3} and three-dimensional systems \cite{Na4Ir3O8}. \\
\indent Most AFM materials on the triangular lattice exhibit  $120^\circ$ long-range magnetic order at low temperatures, such as Ba$_3$CoSb$_2$O$_9$\cite{PRL1_AFM_INS, PRL2_AFM_INS, NC1, NC2Majie}. Exceptions include the organic materials $\kappa$-(BEDT-TTF)$_2$Cu$_2$(CN)$_3$ \cite{ETsalt03} and EtMe$_3$Sb[Pd(dmit)$_2]_2$ \cite{dmit-nc11, NC12gfactor, dmit-science} (we denote it as dmit in later discussion), which exhibit no long-range magnetic order at very low temperatures according to nuclear magnetic resonance (NMR) experiments \cite{ETsalt03, dmit-NMR1, dmit-NMR2} and are considered as QSL candidates. Although these materials are electrically insulating, the low temperature magnetic specific heat is linear in temperature \cite{dmit-nc11}, indicating the existence of plenty of gapless excitations. Thermal transport experiment about dmit indicates that the gapless excitations are mobile since the thermal conductivity is finite when extrapolating to zero temperature \cite{dmit-science} (the existence of residue thermal conductivity at zero temperature is still under debate\cite{LiShiyan2019, PhysRevX19, JPSJ19, JPSJcomment20, JPSJreplycomment, PRBYesAndNo20}). These experiments seem to indicate that the materials are $U(1)$ QSLs containing a finite spinon fermi surface.  On the theoretical side, the ring-exchange interaction ($J_r$ term)\cite{Motrunich} was proposed to stabilize the QSL phases. It was claimed that a $d+id$ chiral spin liquid phase appears at nonzero $J_r$ when the 120$^\circ$ order is suppressed, and an $U(1)$ QSL appears at larger $J_r$. Similar results were obtained later in variational Monte Carlo calculations\cite{XuCenke}.  

In the present work, we study the effect of DM interactions (which are relevant for dmit) on the triangular lattice antiferromagnets, focusing on the thermal transport properties of the  $U(1)$ QSL phase and its stability against magnetic orders. When weak DM interaction is turned on, the spin-up and spin-down spinons feel staggered fluxes in the triangles. Consequently, the shape of the fermi surface and the density of states (DOS) at the fermi level are changed, which affects the longitudinal thermal conductance. Furthermore,  in-plane $\pmb D_{ij}$ components yield nonzero thermal-Hall conductance. Our results qualitatively agree with the thermal transport experiments in the candidate material dmit\cite{dmit-science}. When the ring-exchange is weak or negative, a 120$^\circ$ ordered phase (AMF-I) is observed which is proximate to a $U(1)$ Dirac QSL. In the intermediate region, the DM-interaction caused fermi surface nesting drives the system into another ordered phase (AFM-II). The two AFM phases are distinguished by their static structure factor and their excitation spectrum. The neutron scattering spectrum suggests that Ba$_3$CoSb$_2$O$_9$ belongs to AFM-I. However, the predicted AFM-II phase which appears at finite DM and ring-exchange needs to be identified experimentally.

{\it The Model}. We focus on the following model
\beq\label{JDR}
H=J\sum_{\langle i,j\rangle }\pmb {S_i}\!\cdot\!\pmb {S_j}+D\sum_{\langle i,j \rangle }\pmb {S_i}\!\times\!\pmb {S_j}\!\cdot\!\hat{\pmb D}_{ij}+J_r\sum_{\includegraphics[scale=0.03]{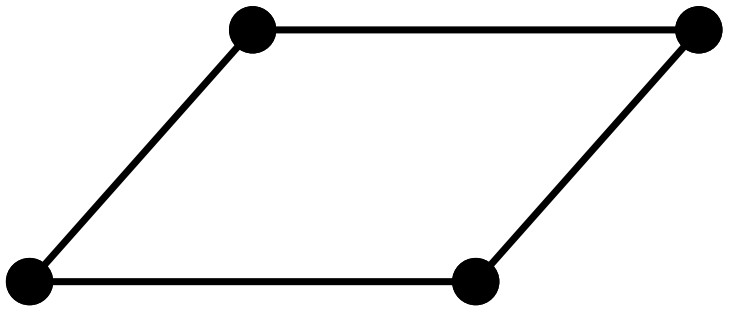}} H_{r}.
\eeq
Among the three kinds of interactions, the $J$-term is the Heisenberg exchange, the $D$-term is the DM interaction, and the $J_r$-term is the ring-exchange\cite{Motrunich} with $H_r$ given by $H_r = (\pmb{S}_i\cdot \pmb{S}_j)(\pmb{S}_k\cdot \pmb{S}_l)+(\pmb{S}_j\cdot \pmb{S}_k)(\pmb{S}_l\cdot \pmb{S}_i)-(\pmb{S}_i\cdot \pmb{S}_k)(\pmb{S}_j\cdot \pmb{S}_l)$ for the parallelogram $i, j, k, l$. $\sum_{\includegraphics[scale=0.03]{tri.jpg}}$ means summing all of the minimal parallelograms. 

\begin{figure}[t]
\includegraphics[width=8.6cm]{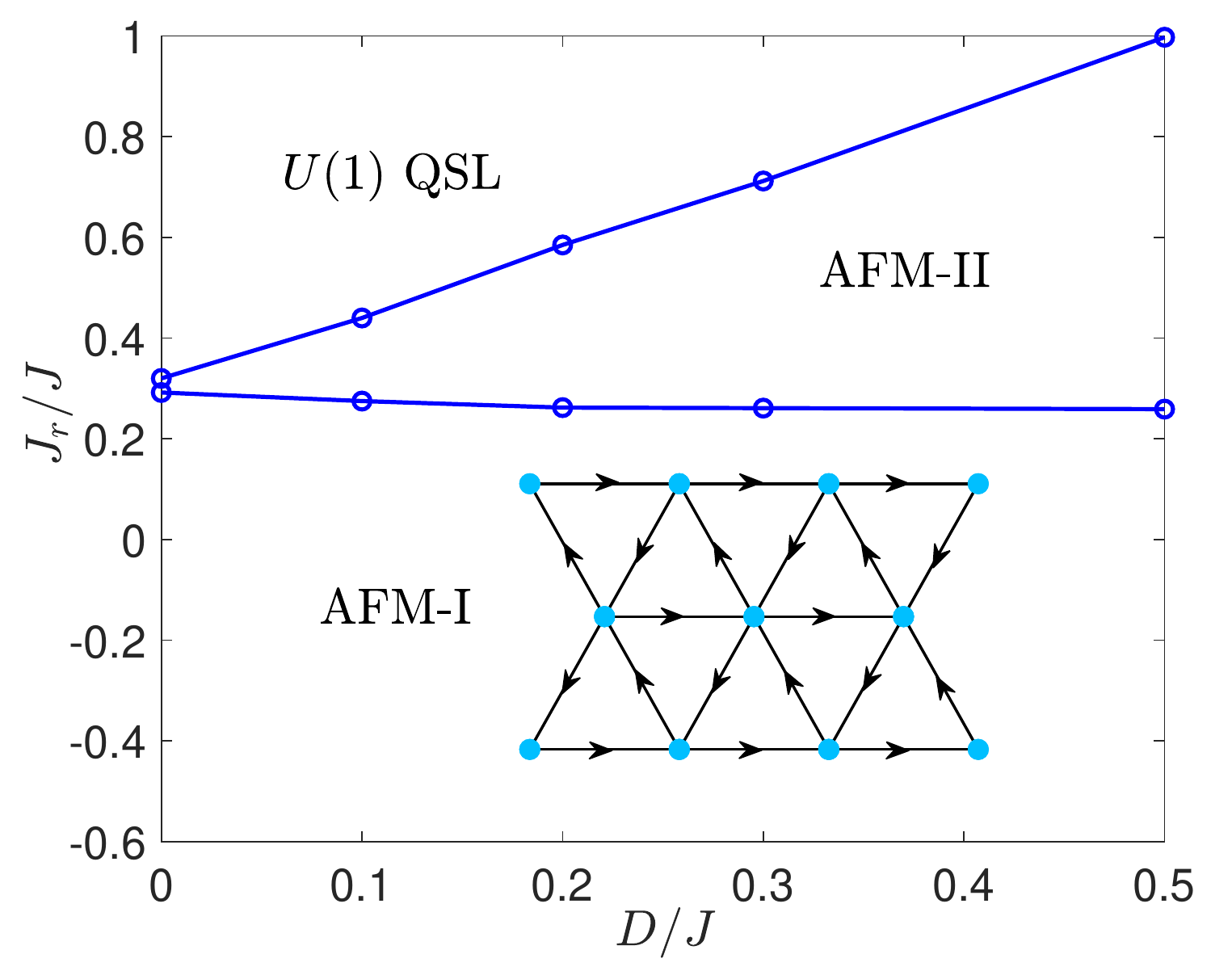}
\caption{Phase diagram of the model (\ref{JDR}), which includes two magnetically ordered phases and one $U(1)$ QSL phase. The $U(1)$ QSL has a finite spinon fermi-surface, while the AFM-I and AFM-II phase exhibit 120$^\circ$ order and approximate 120$^\circ$ order respectively, and are separated by a sharp first order phase transition. The inset illustrates the pattern of the DM interaction, where the vector $\hat{\pmb D}_{ij}$ is along $\hat{\pmb z}$/$-\hat{\pmb z}$ direction if the bond $i\to j$ is along/against the arrows. 
}\label{fig.PhaseDiagram}
 \end{figure}

The $J$ and $J_r$ terms originate from the perturbation theory of the Hubbard model and can be enhanced when the system is close to the Mott transition. The electron spin-orbit coupling yields the DM interaction which only exists when the system has no bond-centered inversion symmetry. For instance, the dimerized [Pd(dmit)$_2]_2${-} anions in  dmit form a segregated stacking structure in a crystal lattice with space group C2/c\cite{dmit-nc11}, where the bond-centered inversion symmetry is absent. The direction of the $\hat{\pmb D}_{ij}$ vector is perpendicular to the $ij$-bond. We first consider the case that the $\hat{\pmb D}_{ij}$ vector is perpendicular to the lattice plane such that the $S_z$ is conserved (see the inset of Fig.\ref{fig.PhaseDiagram}). The effects of in-plane components of the $\hat{\pmb D}_{ij}$ vectors will be discussed later. 

{\it The method.} We study the model (\ref{JDR}) with variational  Monte Carlo. Firstly, we introduce the fermionic parton representation $C_i^\dag= ( c_{i\up}^\dag, c_{i\dn}^\dag  )$ with local particle number constraints $C_i^\dag C_i=1$. In this representation, the the spin operators are written as $S_i^\alpha=C_i^\dagger\frac{\sigma^\alpha}{2}C_i=\bar C_i^\dagger\frac{\sigma^\alpha}{2}\bar C_i$, where $\bar C_i= ( c_{i\dn}^\dag,  -c_{i\up}^\dag ) ^T$ is the particle-hole partner of $C_i$.  The spin operators are invariant under $SU(2)$ mixing of $C_i$ and $\bar C_i$, revealing a $SU(2)$ gauge symmetry \cite{AndersonGauge1988} of the fermion representation. 

Next we decouple the spin Hamiltonian (\ref{JDR}) into a non-interacting fermionic Hamiltonian [see Sec.~S1 of the supplemental materials(SM) for details],
\beq\label{Hmf}
H_{\rm mf} &=&\! \sum_{\langle i,j\rangle}\! \left[ t^{0}_{ij} C_i^\dag C_j + \pmb t_{ij}\!\cdot\! C_i^\dag\pmb \sigma C_j +  \eta^{0}_{ij} C_i^\dag \bar C_j + \pmb \eta_{ij}\!\cdot\! C_i^\dag\pmb \sigma \bar C_j \right. \nonumber \\
&&\left. + {\rm h.c.} \right] 
+ \sum_i (\pmb \lambda \cdot \hat {\pmb \Lambda}_i + \pmb M_i\cdot C_i^\dag {\pmb\sigma\over2}C_i) +{\rm const}, 
\eeq
with $t^0_{ij},\pmb t_{ij}\parallel \hat{\pmb D}_{ij}$ the hopping parameters, $\eta^0_{ij},\pmb \eta_{ij}\parallel \hat{\pmb D}_{ij}$ the paring parameters, $\pmb\lambda$ the Lagrangian multipliers imposing the $SU(2)$ gauge invariance ($\hat {\pmb \Lambda}_i$ are the generators of the $SU(2)$ gauge group),  and $\pmb M_i$ the background field which induces a static magnetic long-range order. To include possible anisotropy in the spin wave function, we introduce a single parameter $\delta$ to denote the relative difference in the amplitudes of parameters between the strong and weak bonds. 
 $\delta\neq0$ indicates the existence of valence bond solid (VBS) order\cite{TaS2_PRL,Song_2020,Song_2019} or nematic order. The state with vanishing magnetic or VBS orders is a QSL. Specially, for $U(1)$ QSLs we can set $\eta^{0}_{ij}=0, \pmb \eta_{ij}=0$.

Then we adopt Gutzwiller projected states $|\Psi(x)\rangle=P_G|G\rangle_{\rm mf}$ as trial wave functions, where $|G\rangle_{\rm mf}$ is the ground state of $H_{\rm mf}$, $P_G$ stands for the Gutzwiller projection that enforces the local particle number constrain, and $x=(t^{0}_{ij}, \pmb t_{ij}, \eta^{0}_{ij}, \pmb \eta_{ij}, \pmb M_i, \pmb\lambda,  \delta)$ are the variational parameters. Different types of ansatz,  including various $Z_2$ QSLs, $U(1)$ QSLs, and 120$^\circ$ ordered states have been considered as the initial states in the variation process. The optimal values of $x$ are determined by minimizing the trial energy $E(x)= \langle \Psi | H |\Psi\rangle/\langle \Psi |\Psi\rangle$.  We adopt a torus geometry with $L_a=L_b$ ($L_{a,b}$ are the lengths along the $\pmb a$-/$\pmb b$-axes where $\pmb a\cdot\pmb b=-{1\over2}$) up to $L_a=L_b=12$.

The phase diagram turns out to be very simple. As shown in Fig.\ref{fig.PhaseDiagram}, only one $U(1)$ QSL phase plus two magnetically ordered phases (labeled as AFM-I and AFM-II) are found. All of the phases transitions are of first order. 

 We have considered potential VBS order or nematic order. It turns out that the ground state may contain certain VBS\cite{TaS2_PRL, Song_2020, Song_2019} or nematic order if the system has geometry $L_a\neq L_b$. But if $L_a = L_b$, neither the VBS nor nematic order is energetically favored (see SM for details). We infer that in the thermodynamic limit the system contains not VBS or nematic order.

{\it The $U(1)$ QSL phase.} The disordered $U(1)$ QSL phase with a finite fermi surface appears near the region $(J_r - 1.4D)/J > 0.3$. Specially, when $D=0$ the system is rotation invariant, thus the triplet hopping terms vanish $t^{x,y,z}_{ij}=0$ and the singlet parameters $t^{0}_{ij}$ is a real number.

\begin{figure*}[t]
\includegraphics[width=4.3cm]{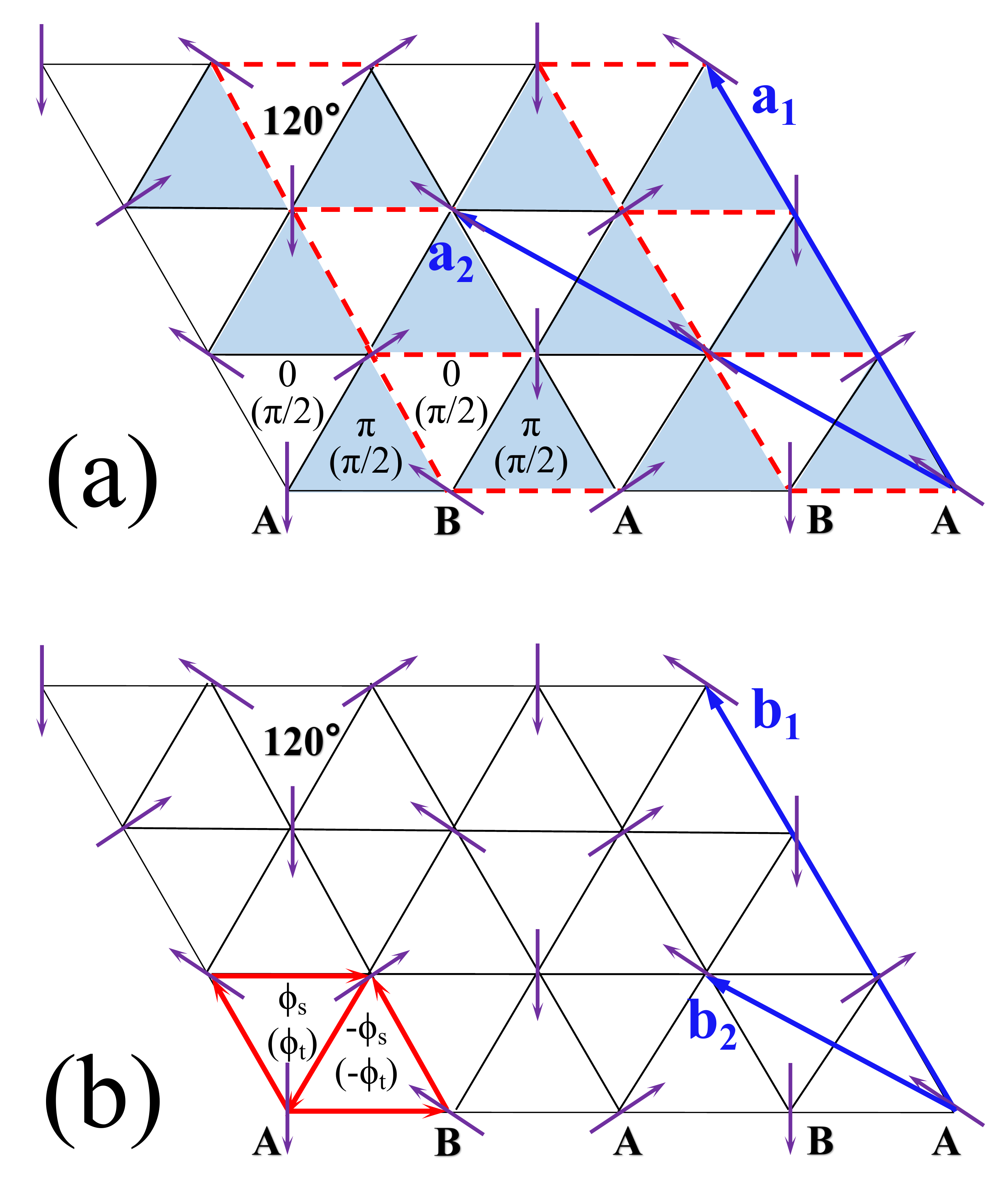}\label{fig:unitcell_pi_stagger}\!\!\!
\includegraphics[width=7.0cm]{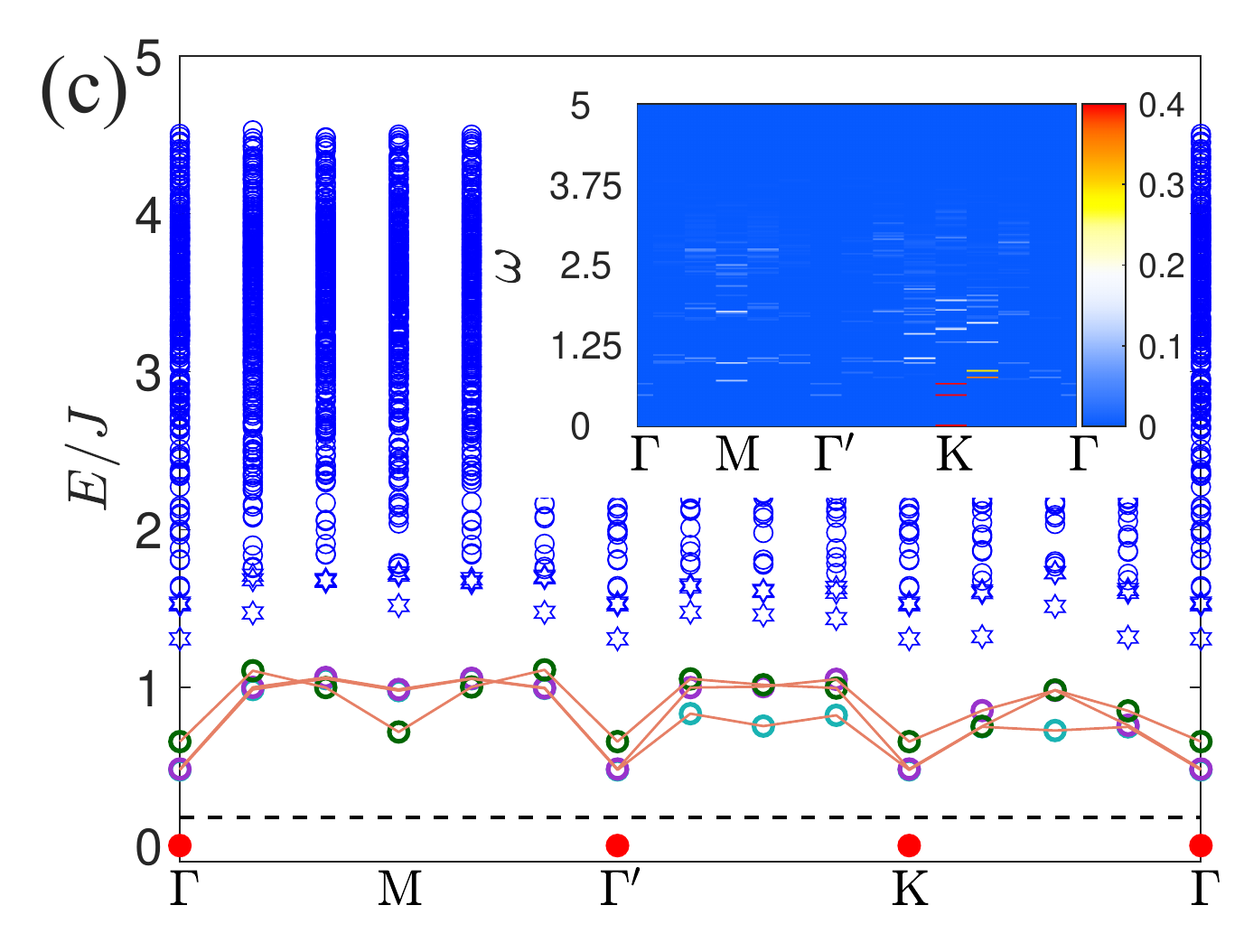}\!\!\!\!
\includegraphics[width=7.0cm]{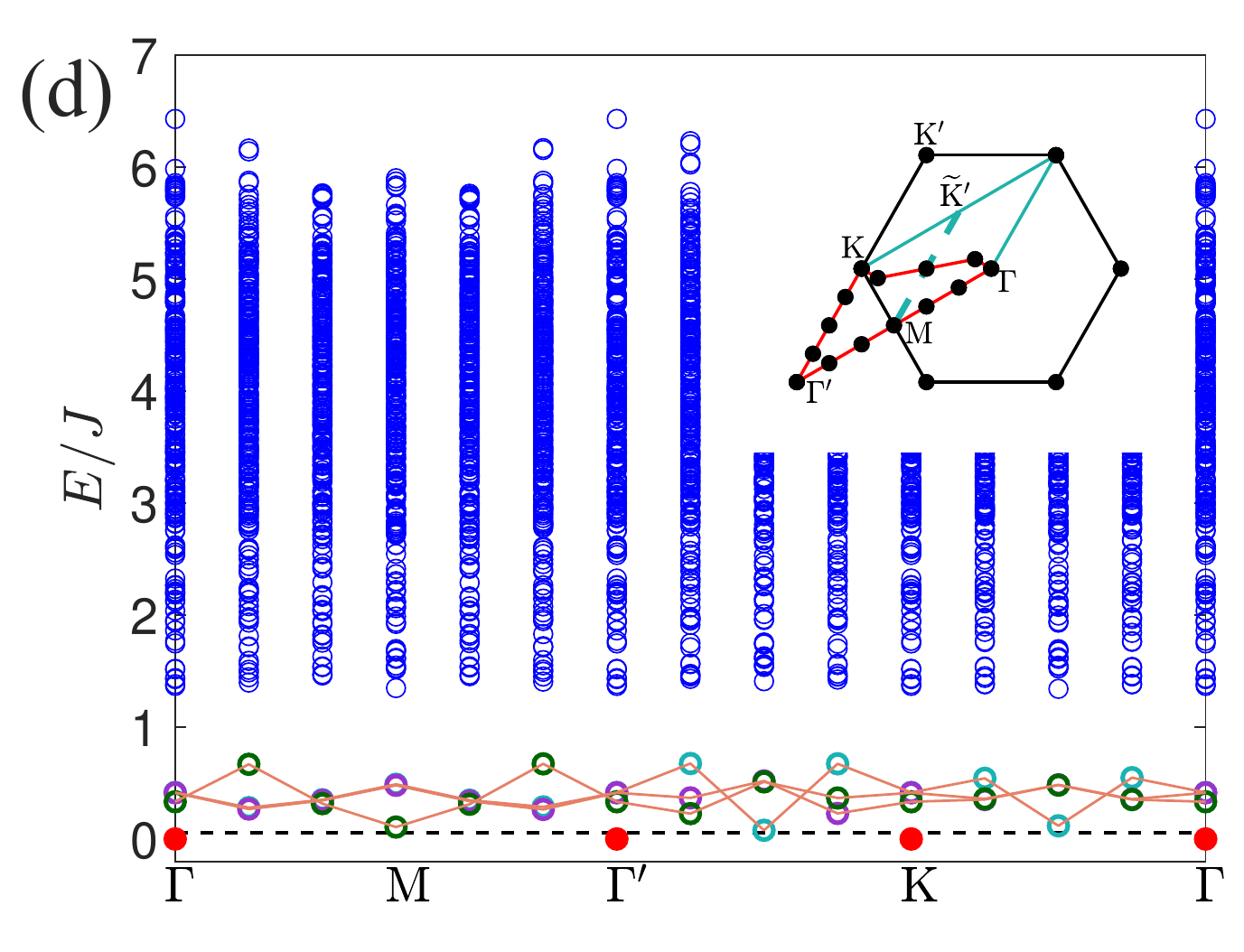}\!\!\!\!
\caption{(a)\&(b),The unit cells of the AFM-I and AFM-II phases, respectively. In AFM-I, the singlet  hopping terms contain 0/$\pi$ flux on the white/dark triangles, and the triplet hopping terms contain uniform ${\pi\over2}$-flux. In AFM-II, the singlet/triplet hopping terms see staggered $\phi_s$-flux/$\phi_t$-flux. 
(c)\&(d), The one-magnon bands and continuum spectra (in a system with 72 sites) for AFM-I (${D\over J}=0, {J_r\over J}=0$) and AFM-II (${D\over J}=0.1, {J_r\over J}=0.35$), respectively.  The dashed line is the energy of the ground state before the `renormalization', while the red dot stands for the `renormalized' one. The inset of (c) shows the dynamic structure factor of the Heisenberg model, and the inset of (d) illustrates the Brillouin Zone and the path of the dispersion curves.
}\label{fig:magnon}
 \end{figure*}

When the DM interaction is turned on, the singlet hopping $t^{0}_{ij}$ becomes complex and the triplet hopping parameters $t^{x,y,z}_{ij}$ acquire finite (complex) values. When $\hat{\pmb D}_{ij}\parallel \hat{\pmb z}$, $S_z$ is conserved and the spin-up and spin-down fermions see staggered fluxes in the triangles [see Fig.~\ref{fig:magnon}(b) for illustration of the flux pattern]. The time-reversal symmetry $T$ is not broken because the state can be transformed into its original form by a $SU(2)$ gauge transformation after the physical $T$ operations. However, the spatial inversion symmetry $P$ is indeed broken. 

\begin{figure}
\includegraphics[width=4.2cm]{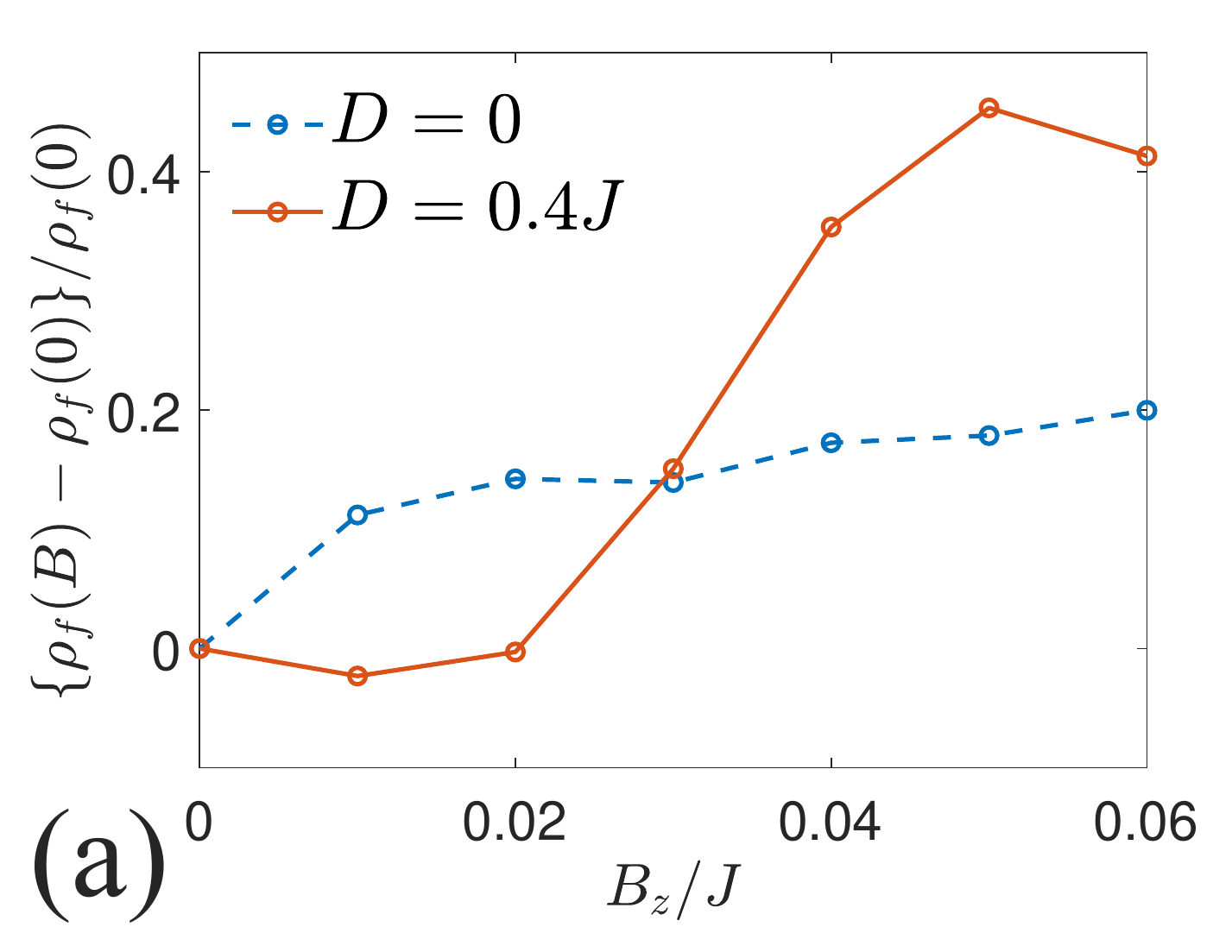}\ \ 
\includegraphics[width=4.2cm]{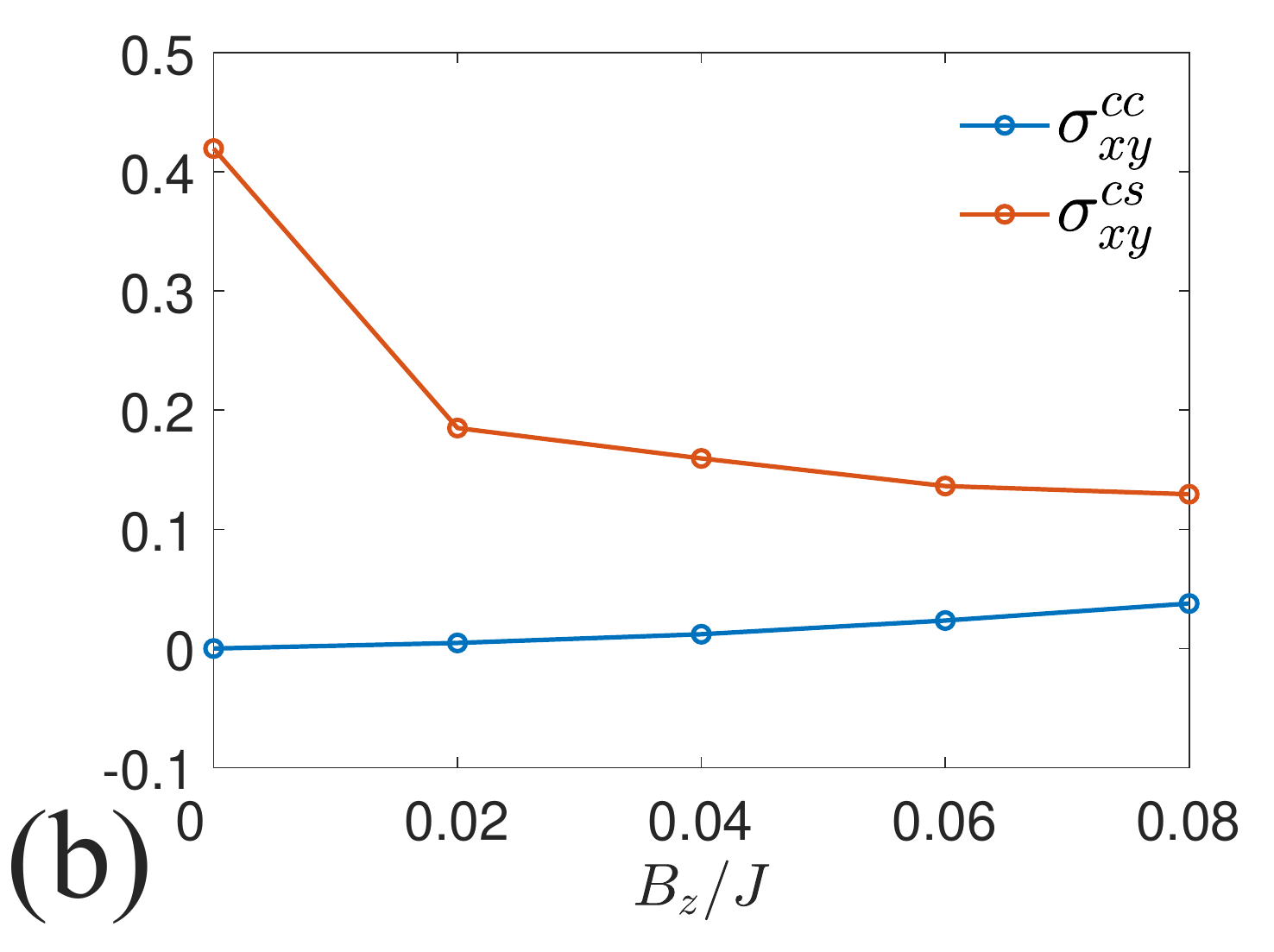}
\includegraphics[width=4.0cm]{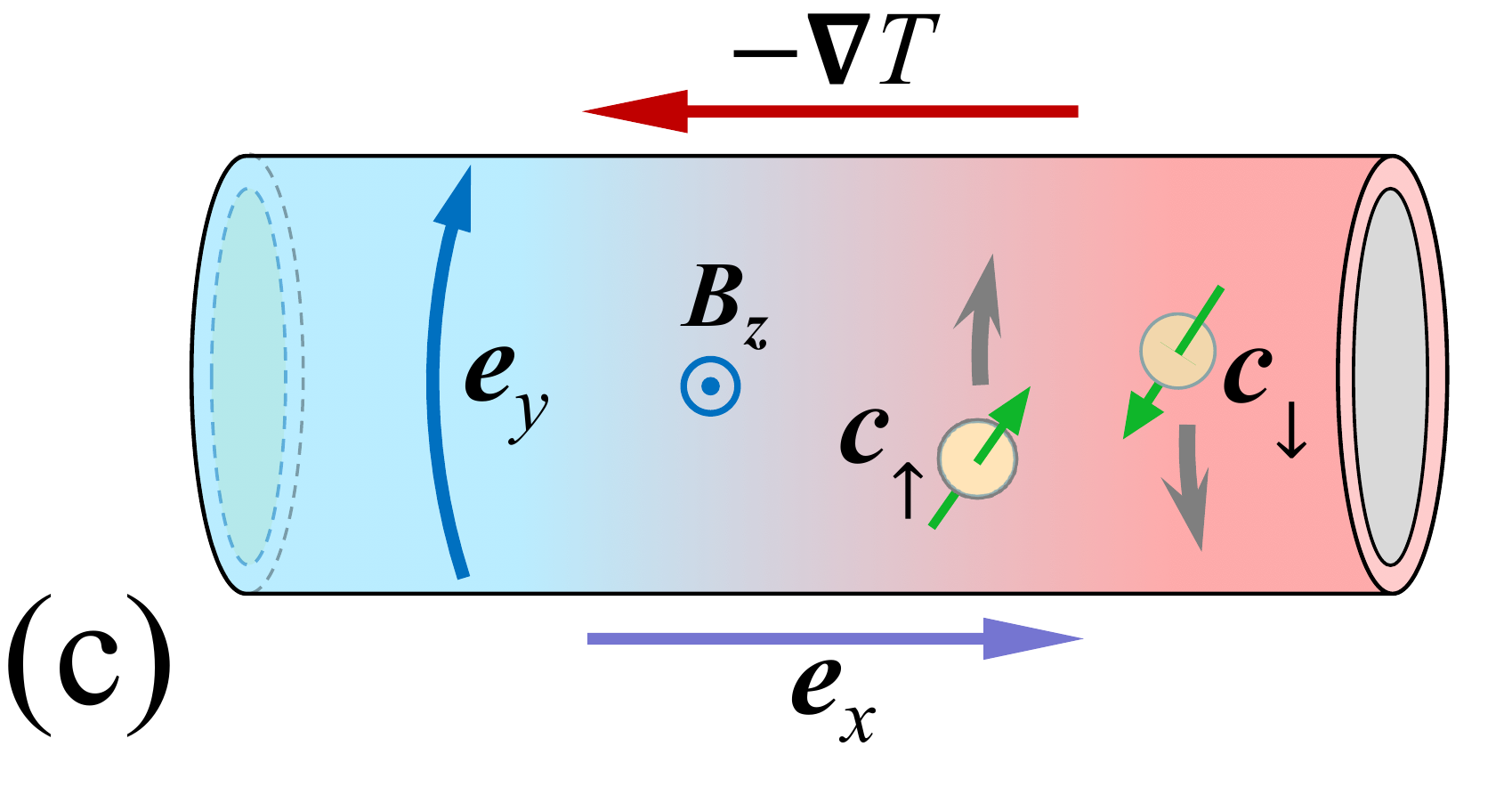}\ \ \ 
\includegraphics[width=4.1cm]{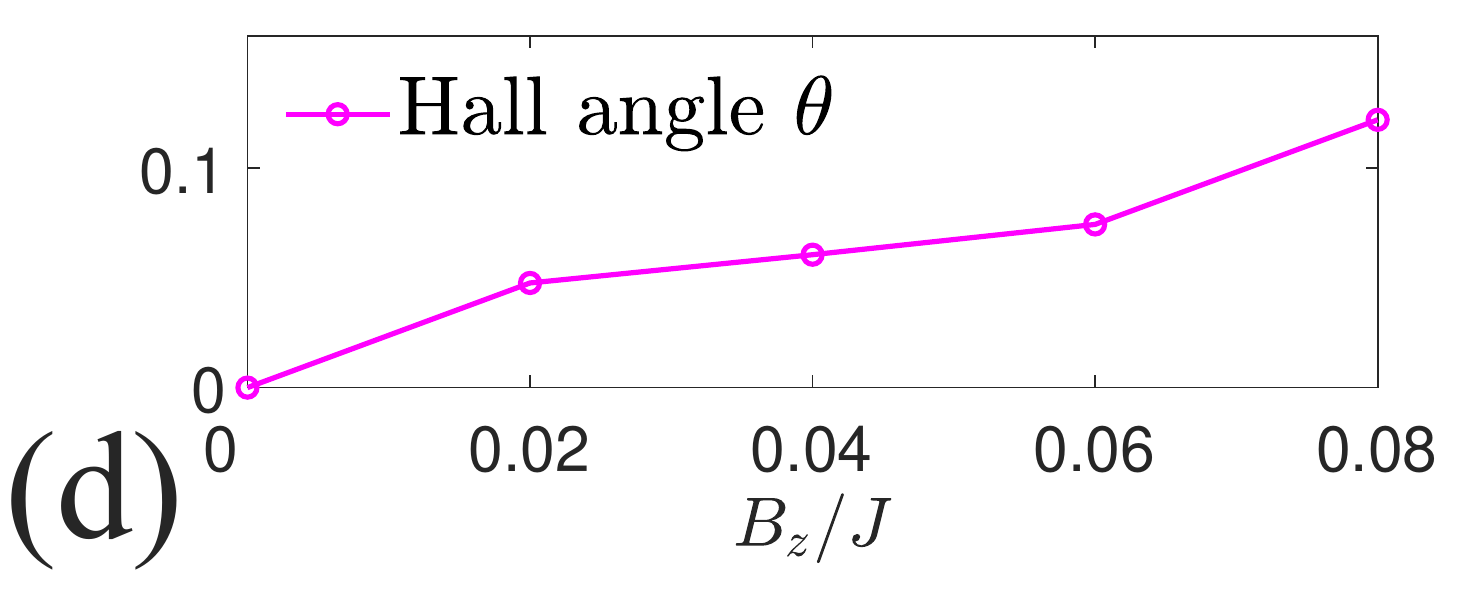}
\caption{(a) DOS at the fermi surface as a function of magnetic field $B_z$. (b) Mean field gauge charge-charge and charge-spin Hall conductances $\sigma^{cc}_{xy}$ and $\sigma^{cs}_{xy}$ as functions of $B_z$ (here we assume that the angle between $\pmb D_{ij}$ and $\hat{\pmb z}$ is $\pi\over16$). (c) Cartoon picture of the thermal Hall conductance, where $e_x$ and $e_y$ are the emergent gauge `electric' field. (d) The calculated thermal Hall angle as a function of $B_z$.
}\label{fig:Hall}
 \end{figure}

The triplet hopping terms split the degeneracy between spin-up and spin-down spinons. Especially, the staggered flux results in deformation of the fermi surfaces (the two fermi surfaces have different shapes, see Fig.S1 in Sec.2 of the SM).  As external magnetic field $B_z$ is turned on, the $c_\up$ and $c_\dn$ bands shift in opposite directions and resultantly the density of states (DOS) $\rho_f$ at the fermi level is dependent on the field strength $B_z$. The DOS $\rho_f$ affects the thermodynamic properties of the spin system. For instance, the specific heat $c_v = {\pi\over3} k_B^2 T \rho_f(B_z)$ (here we ignore the contribution from gauge photons and will address it later) is proportional to $\rho_f$, where $k_B$ is the Boltzmann constant and $T$ is the temperature. Similarly, the longitudinal thermal conductance $\kappa_{xx}$ depends on the DOS of the spinon fermi surface via $\kappa_{xx}={1\over3}v_F^2\tau c_v = {\pi\over9}k_B^2v_F^2\tau T \rho_f(B_z)$, namely 
\beq\label{kappa_xx}
\kappa_{xx}/T \sim \rho_f(B_z), 
\eeq
where $v_F$ is the fermi velocity and $\tau$ is the mean free time between spinon collisions. Fig.\ref{fig:Hall}(a) illustrates the DOS $\rho_f(B_z)$ as a function of $B_z$. It can be seen that the $B_z$-dependence of $\kappa_{xx}$ is changed by the DM interaction.

Another important effect of the DM interaction is that it may generate nonzero thermal Hall effect\cite{Nagaosa10_thermal,ChenGang1}. When the $\hat{\pmb D}_{ij}$ vector contains in-plane component, the $c_\up$ and $c_\dn$ spinons couple to each other, consequently the spinons obtain non-trivial Berry phase in momentum space. The nontrivial Berry phase results in nonzero gauge charge-charge Hall conductance $\sigma^{cc}_{xy}$ and gauge charge-spin Hall conductance $\sigma^{cs}_{xy}$ at mean-field level.

Suppose the system has a temperature gradience $\nabla T$ along $\hat{\pmb x}$-direction, then a longitudinal spinon current $ j_0^c = -\zeta_{xx}{\partial T\over\partial x }$ is generated along $-\hat{\pmb x}$ direction at mean field level, where $\zeta_{xx}$ is the longitudinal thermal charge conductance [$\zeta_{xx}$ is also proportional to $\rho_f(B_z)$]. Under external magnetic field $B_z$, $\nabla T$ may also drive a transverse spinon current $ j_1^c = -\zeta_{yx}{\partial T\over\partial x }=\zeta_{xy}{\partial T\over\partial x }$ flowing along $\hat{\pmb y}$ direction with $\zeta_{xy}$ the thermal-charge Hall conductance. However, the single occupancy condition forbids the net spinon density and current fluctuations. The Ioffe-Larkin rule\cite{IoffeLarkin89dimer, LeeNagaosa92} suggests that the effect of the local particle constraint can be analyzed under the mean-field framework. Essentially, an internal gauge `electric' field $\pmb e$ is induced to generate spinon currents to cancel the longitudinal and transverse particle transport.

Considering that both $ e_x$ and $ e_y$ generate longitudinal and transverse spinon current, their values should be determined by the net current-vanishing conditions,
\Beq
 {j}_x^c &=& \sigma^{cc}_{xx} {e}_x + \sigma^{cc}_{xy} {e}_y -  {j}_0^c= 0,  \\ 
 {j}_y^c &=& \sigma^{cc}_{yx} {e}_x + \sigma^{cc}_{yy} {e}_y + {j}_1^c= 0,
\Eeq
with $\sigma^{cc}_{yx}=-\sigma^{cc}_{xy}$ and $\sigma^{cc}_{yy}=\sigma^{cc}_{xx}$. The solution is given by 
$
{e}_x = \frac{\sigma^{cc}_{xx}{j}_0^c + \sigma^{cc}_{xy}{j}_1^c }{ (\sigma^{cc}_{xx})^2+ (\sigma^{cc}_{xy})^2} ,$ and  
${e}_y =  \frac{\sigma^{cc}_{xy} {j}_0^c - \sigma^{cc}_{xx}{j}_1^c }{ (\sigma^{cc}_{xx})^2 + (\sigma^{cc}_{xy})^2}.
$
On the other hand, owing to the gauge charge-spin Hall effect, the $e_x$ generates a finite spin current $j^{s_z}_y = -\sigma^{cs}_{xy} e_x$ along $\pmb y$-direction. Since Zeeman effect results in an average energy difference $\delta E = - g\mu_B B_z$ between the $c_\up$ and $c_\dn$ spinons, the spin current carries a finite energy current,  $ j^e_{y} = -\sigma^{cs}_{xy} {e}_x\delta E = g\mu_B B_z\sigma^{cs}_{xy}\frac{\sigma^{cc}_{xy}\zeta_{xy} -\sigma^{cc}_{xx}\zeta_{xx} }{(\sigma^{cc}_{xx})^2+(\sigma^{cc}_{xy})^2} \partial_xT$. Generally $\sigma^{cc}_{xy}\ll \sigma^{cc}_{xx}$ and $\zeta_{xy}\ll \zeta_{xx}$, so we have
\beq
\kappa_{xy}=-j^e_y/\partial_xT   \simeq  g\mu_B B_z\sigma^{cs}_{xy}{\zeta_{xx}\over\sigma^{cc}_{xx}} \propto g\mu_B B_z\sigma^{cs}_{xy}T, 
\eeq
where we have used the relation ${\zeta_{xx}\over\sigma^{cc}_{xx}}=S$ where $S\propto T$ is the Seebeck coefficient (see Sec. S5 in the SM for details). 
The mutual Hall conductance $\sigma^{cs}_{xy}$ results from the Berry phase of the spinons below the fermi energy. Furthermore, magnetic field changes the size of the fermi-surface, so $\sigma^{cs}_{xy}$ also depends on the field intensity $B_z$ [see Fig.\ref{fig:Hall} (b)].  From above analysis we obtain the $B_z$ dependence of the thermal Hall angle
$$
\theta\sim \tan \theta = {\kappa_{xy}\over\kappa_{xx}}  \propto  {B_z\sigma^{cs}_{xy}\over\rho_f(B_z)},
$$ 
which is illustrated in Fig.\ref{fig:Hall}(d) for $D/J=0.4$.

If the spinon bands are fully gapped, $\kappa_{xy}/T$ will be quantized at extremely low temperatures\cite{Kitaev06, GongSS2016,Nagaosa10_thermal,ChenGang1}. However, since the singlet hopping terms are dominating, the fermi surface cannot be gapped out unless $D$ or $B_z$ is very strong. But the resultant state is either magnetically ordered or full polarized. Therefore, quantized thermal Hall effect does not occur in the present model.


{\it Magnetically ordered phases.} When the ring-exchange interaction is weak or when the DM interaction is strong, we obtain two different AFM ordered phases labeled as AFM-I and AFM-II respectively. The ansatz of the ordered states are given in (\ref{Hmf}), where $\pmb M_i$ is the background field inducing the magnetic order, the rest spin liquid terms govern the quantum fluctuations. In the AFM-I phase, the singlet hopping terms contain alternating 0- and $\pi$-flux in the triangles (which yields a $U(1)$ Dirac QSL if there are no triplet hopping terms), and the triplet hopping terms contain uniform ${\pi\over2}$-flux [which gaps out the Dirac QSL into a quantum spin Hall state (QSH), see Fig.\ref{fig:magnon}(a) for illustration]; the AFM-II phase is proximate to the $U(1)$ QSL in Fig.\ref{fig.PhaseDiagram} where the flux pattern of the singlet and triplet hopping terms are shown in Fig.\ref{fig:magnon}(b). 

The ordering mechanism of the two AFM phases are quite different. The AFM-I is resulting from monopole proliferation in a QSH state. As mentioned previously, the QSH state is a massive $U(1)$ Dirac QSL whose mass comes from the DM interaction or spontaneous generation\cite{Song_2020, Song_2019}. Since the monopoles carry momentum $\pmb K$, their proliferation yields 120$^\circ$ order in AFM-I. 
On the other hand, the order in AFM-II results from the nesting of the fermi-surface when the triplet hopping is finite. The nesting only occurs at the lower fermi surface (see SM) and the nesting momentum slightly deviates from $\pmb K$. As a result, the AFM-II is not precisely $120^\circ$ ordered. The difference in the ordering momentum is experimentally observable and can be used to distinguish the two AFM phases. 
However, in the ansatz for AFM-II, we still approximately adopt the 120$^\circ$ order. Even though, the sharp first order phase transition between the two AFM phases can be observed from the discontinuity in the order parameters.  
 
Now we  study the characters of the AFM phases in the excitation spectra.
The low-energy excitations in magnetically ordered states are magnons and the fermionic spinons should be confined in the low energy limit. However, the `confinement' of the spinons is not directly reflected in the VMC approach. To partially solve this problem, we focus on the Hilbert subspace spanned by the family of Gutzwiller projected states, each of which contains one excited spinon plus one hole with the total momentum fixed. Then we diagonalize the Hamiltonian (\ref{JDR}) in this subspace to obtain the dispersion of the `renormalized' excitations \cite{YangFanLiTao2010, YangFanLiTao2011, NPhys14_GutzSpectrum}. 
In the low-energy section of the resultant spectrum, the spinon and hole are bound to form an `exiton'-like quasiparticle. The dispersion of this bound quasiparticle forms several energy bands which are separated from the higher-energy continuum by a binding energy gap. As shown in Figs.~\ref{fig:magnon}(c) and \ref{fig:magnon}(d) for the AFM-I and AFM-II phases, the average number of modes for the quasiparticle is one per site. Hence these modes can be identified as a single magnon, namely the goldstone modes for $\pmb D^{x,y}_{ij}=0$ (the gap between the ground state and the magnon bands is due to finite size effect). 
%

The ratio between the binding energy gap (i.e gap between the magnon bands and the higher-energy continuum) and the magnon band width reflects how tight the magnon is formed. This ratio in AFM-I phase is smaller than that in AFM-II phase (see Fig.\ref{fig:magnon}). 
Furthermore, the AFM-I have a lot of `mid-gap' modes (marked as the hexagrams) lying in the gap between the magnon bands and the continuum. These modes can be considered as the second quasi-particle like bound states formed by the excited spinon and the hole, which constitute another three bands (two of the bands are adjacent to the continuum). No `mid-gap' modes are seen in AFM-II.


{\it Comparing with experiments.}  The nonzero thermal conductance $\kappa_{xx}$ in dmit indicates that there are highly mobile low-energy quasiparticles with a large density of states, which is considered the evidence for the existence of spinon fermi-surface. Especially,  at very low temperature, $\kappa_{xx}$ first decreases and then increases with increasing $B_z$. This field dependence can be qualitatively interpreted by the change of DOS of the fermi surface, whose field dependence $\rho_f(B_z)$ is shown in Fig.\ref{fig:Hall}(a) with $D/J=0.4$. This suggests that dmit has nonzero DM interactions, which is allowed by the symmetry of the material. Furthermore, at low temperatures dmit shows a weak thermal Hall conductance. The thermal Hall angle is roughly linear in $B_z$. From our theoretical analysis,  the thermal Hall conductance of the QSL is non-vanishing only when the spinons have a finite spin-orbit coupling such that the Berry phase of the spoinons is nonzero in momentum space. For dmit, one possibility is that the $\pmb D_{ij}$ vectors have nonzero $x,y$-components.

However, quantum field theory predicts that for a $U(1)$ QSL with spinon fermi surface the gauge photons modify the linear $T$ dependence of $c_v\sim T$ into $c_v\sim T^{2/3}$, which is inconsistent with dmit.  A possible solution is that the spinons pair up to form a $Z_2$ QSL \cite{YaoHonngMaissam} such that the gauge photons are Higgsed, while the spinon fermi surface still survives as inversion symmetry is absent. Our VMC results indicate that $Z_2$ QSL phase is energetically unstable in the $J$-$J_r$-$D$ model. One possibility is that more complicated spin-orbital coupling interactions which lower the symmetry of the system may provide the force to glue the spinon pairing. We leave these issues for future study.  

On the other hand, the $120^\circ$ ordered  triangular magnet Ba$_3$CoSb$_2$O$_9$ is approximately described by the Heisenberg model, and falls in the AFM-I phase.  Inelastic neutron scattering\cite{NC1, Majie_2016, Radu_2020} indicates that the shape (roton-like minima) and the energy range ($\sim 1J$) of the magnon bands around the $M$ point deviate from linear spin wave theory,  but these features are in quantitative agreements with our results [see the spectrum and dynamic structure factor\cite{DSF1D_Sorella_18,DSF_PRX19} shown in Fig.\ref{fig:magnon}(c)]. Especially, the large weight of the structural continuum correspond to the `mid-gap' modes and the higher-energy spinon-hole continuum. 

{\it Conclusions.}  In summary, we have studied the effect of DM interaction to the $U(1)$ QSL on the triangular lattice. The DM interaction introduces staggered fluxes for the spinons, which changes the shape of the fermi surface and affects the longitudinal thermal conductance. Especially when $\hat{\pmb D}_{ij}$ have nonzero in-plane components, then the spinons acquire spin-orbital coupling and consequently the system exhibits a nonzero thermal Hall effect. These results are qualitatively consistent with the experimental result of EtMe$_3$Sb[Pd(dmit)$_2]_2$.

We predict a new phase, i.e. the AFM-II appearing between the $U(1)$ QSL and the AFM-I. The magnetic order in AFM-II originates from fermi-surface nesting and is different from that in AFM-I which results from monopole proliferation.  The two AFM phases can be distinguished by their static structure factors and excitation spectra.

Many other ansatz, including the gapped $Z_2$ QSL, Kalmayer-Laughlin chiral QSL, $d+id$-wave chiral QSL, nodal $d$-wave QSL and $U(1)$ Dirac QSL, have been studied but none of them are energetically stable in the present model. While some of these phases, such as the  Kalmayer-Laughlin chiral QSL has been realized in different models\cite{GongSS2016}, the realization of the remaining QSL phases is challenging and deserves future study.

\textit{Acknowledgement} -- We thank J.-C. Wang, T. Li, C. Zhang, G. Chen, R. Yu and Y.-C. He for valuable discussions. This work is supported by the Ministry of Science and Technology of China (Grant No. 2016YFA0300504), the NSF of China (Grants No. 11574392 and No. 11974421), and the Fundamental Research Funds for the Central Universities and the Research Funds of Renmin University of China (No. 19XNLG11).

\bibliography{U1DM}

\end{document}